\documentclass[twocolumn,superscriptaddress,aps,prb]{revtex4-1}
\usepackage{graphicx}% Include figure files
\usepackage{dcolumn}% Align table columns on decimal point
\usepackage{bm}
\usepackage{color,graphicx}
\usepackage{amsmath}
\usepackage{eqnarray}
\usepackage{amssymb}
\usepackage{amsthm}
\usepackage{amsfonts}
\usepackage{braket}
\usepackage[arrowdel]{physics}
\pdfoutput=1 % for ArXiv

\begin{document}

%\title{A model for a leaky dielectric with memory}
\title{Microscopic Modeling of Polarization Dynamics in Leaky Dielectrics: Insights into Ferroelectric-Like Behavior}

\author{Igor Ricardo Filgueira e Silva}
%\email{igor@estudante.ufscar.br}
\affiliation{Departamento de Fisica, Universidade Federal de Sao Carlos, 13565-905, Sao Carlos, SP, Brazil}

\author{Ovidiu Lipan}
\affiliation{Department of Physics, University of Richmond, 28 Westhampton Way, Richmond, Virginia 23173, USA}

\author{Fabian Hartmann}
\affiliation{Julius-Maximilians-Universitat Wurzburg, Physikalisches Institut and Wurzburg-Dresden Cluster of Excellence ct.qmat, Lehrstuhl fur Technische Physik, Am Hubland, 97074 Wurzburg, Deutschland}

\author{Sven H\"{o}fling}
\affiliation{Julius-Maximilians-Universitat Wurzburg, Physikalisches Institut and Wurzburg-Dresden Cluster of Excellence ct.qmat, Lehrstuhl fur Technische Physik, Am Hubland, 97074 Wurzburg, Deutschland}

\author{Victor Lopez-Richard}
%\email{vlopez@df.ufscar.br}
\affiliation{Departamento de Fisica, Universidade Federal de Sao Carlos, 13565-905, Sao Carlos, SP, Brazil}

\begin{abstract}

Based on a microscopic model of nonequilibrium carrier generation in a leaky dielectric, we analytically derive hysteresis loops for the dielectric response of non-polar, non-ferroelectric materials. We demonstrate how complex dielectric responses can emerge solely from the influence of transport processes that depend on energy levels, voltage polarity, and asymmetries in charge transfer rates. By combining Electrochemical Impedance Spectroscopy and voltammetry, we address critical questions related to the microscopic mechanisms in poorly conductive systems dominated by displacement currents. The impedance analysis, extended to higher-order harmonics, provides deeper insights into the dynamic behavior of dielectric materials, emphasizing the need to correlate impedance spectroscopy with dielectric spectroscopy for a thorough understanding of dipole relaxation and transport phenomena. Our approach provides a fully analytical framework that directly correlates microscopic charge dynamics with macroscopic dielectric responses, offering enhanced accuracy and predictive capability for systems dominated by displacement currents.

\end{abstract}

\maketitle

\section{Introduction}

In the study of dielectric and polar materials, several fundamental questions are still relevant, particularly concerning the interplay between dielectric responses and carrier drift under nonequilibrium dynamics~\cite{Chen2022,Wang2024}, specifically in tuning dipole relaxation and leakage effects~\cite{Liu2020a,Wang2023}. Electrochemical impedance spectroscopy (EIS)~\cite{Irvine1990,Barsoukov2018}, dielectric spectroscopy~\cite{Woodward2021,Clark2024}, and voltammetry~\cite{Balke2015,luo2023well} are powerful techniques widely used in material science and nanoscale device characterization to explore these issues. However, traditional approaches to such techniques may hinder the extraction of relevant information about these nonequilibrium processes. 

Polar materials such as perovskites, ferroelectrics, piezoelectrics and dielectrics in general are of significant interest due to their potential applications in sensors~\cite{Cui2022}, actuators~\cite{Liu2023}, memory devices~\cite{Kittl2009}, and photovoltaic cells~\cite{Aqoma2024}. Understanding the microscopic mechanisms in these materials, especially in poorly conductive systems dominated by displacement currents, is crucial for optimizing their performance and expanding their technological applications. This paper aims to address several critical questions, offering deeper insights into these underlying phenomena.

Firstly, we explore how the intermix of dielectric response and carrier drift affects voltammetry and impedance from a microscopic perspective. Understanding these interactions is essential to accurately interpret the data obtained from these techniques. Secondly, we investigate potential mimetic effects arising from the interplay between dielectric response and carrier drift, which might be misinterpreted.

The investigation of ferroelectric-like behavior in leaky dielectrics has long been complicated by experimental artifacts and the difficulty of isolating genuine ferroelectric phenomena. For instance, Scott in Ref. \cite{scott2007ferroelectrics} humorously yet critically highlighted how lossy dielectrics, including even household objects like bananas, can produce hysteresis loops that mimic ferroelectric behavior due to leakage currents. These artifacts often arise from measurement configurations that fail to account for conductivity effects, as emphasized by the presence of cigar-shaped loops instead of saturated, concave hysteresis curves. Vasudevan et al. in Ref. \cite{Vasudevan2017} highlighted how piezoresponse force microscopy can reveal hysteresis loops and remnant polarization states in non-ferroelectric materials, often misinterpreted as evidence of true ferroelectricity. Similarly, Balke et al. in Ref. \cite{Balke2015} and Ganeshkumar et al. in Ref. \cite{ganeshkumar2017decoding} demonstrated that piezoresponse force microscopy and contact Kelvin probe force microscopy can uncover the contributions of electrostatic interactions and charge trapping to polarization dynamics in non-ferroelectric materials.

These studies underscore critical limitations in conventional experimental methodologies, particularly the reliance on macroscopic hysteresis measurements that fail to differentiate between genuine ferroelectric switching and effects from leakage or interface dynamics. By integrating these insights, our model provides a theoretical framework that explicitly incorporates the interplay between charge trapping and leakage currents, allowing for a more accurate interpretation of complex dielectric responses. This approach offers guidance for experimental validation through protocols such as temperature- and frequency-dependent impedance spectroscopy, which are capable of isolating and identifying specific contributions to the observed polarization dynamics.

%Our findings reveal that there are additional nuances to the ferroelectric mimicry explored in Ref.~\citenum{scott2007ferroelectrics}, highlighting the complexity of these interactions. Distinguishing between these effects is crucial to avoid erroneous conclusions about material properties and device behavior~\cite{Esquivel2011,Balke2015,Vasudevan2017,ganeshkumar2017decoding}. 

Furthermore, we examine the differences in parameter interpretation when using a microscopic model versus traditional equivalent circuit representations~\cite{Irvine1990,Li2011,Vadhva2021,Oliveira2020,Ramirez2023}. This comparison highlights the limitations of conventional models and the potential inaccuracies that may arise from their use.

In this context, we examine the implications of using equivalent circuits that may not fully capture the underlying microscopic processes in dielectrics. While equivalent circuits offer a valuable framework for simplifying complex systems and aiding in the interpretation of impedance data~\cite{Bisquert2024a,Bisquert2024b}, projections into stationary circuit elements without a proper microscopic basis can sometimes result in inconsistencies or unexpected behaviors. This highlights the value of incorporating microscopic insights to enhance the accuracy and reliability of equivalent circuit models, while still benefiting from their ability to streamline device analysis and design. Additionally, we discuss how to correlate impedance analysis, dielectric spectroscopy, and higher mode generation with microscopic features and structural parameters. This correlation is vital for developing a comprehensive understanding of the material and device characteristics and we are able to provide it in fully analytical form. Lastly, we identify what might be missing in traditional impedance and voltammetry characterization, with an emphasis on multimode analysis. This approach can uncover additional dimensions of information that are often overlooked in conventional analyses.

By addressing these questions, this paper aims to enhance the  reliability of EIS and voltammetry techniques, providing more precise guidelines for their application in material science and device research. This enhanced understanding is particularly crucial for systems involving dielectrics and leaky dielectrics, such as capacitors, metal-oxide interfaces in transistors, and floating gate memories. In these systems, the dynamic behavior of electric polarization plays a pivotal role, influencing performance and stability.

The generality of the relations embedded in our model, combined with the simplicity of the final analytical expressions, ensures its applicability to a broad class of leaky dielectrics. This versatility stems from the ubiquitous presence of charge traps at defects or interfaces, which serve as the primary mechanisms influencing dielectric responses in semiconductors and dielectric materials alike.

\section{Results}
Defects and material properties play critical roles in shaping polarization dynamics in dielectric systems \cite{Esquivel2011}, directly influencing the assumptions of the proposed model. Point defects, such as oxygen vacancies and interstitial atoms, serve as charge traps that impact conductivity and polarization through trapping and detrapping mechanisms, with their energy levels within the bandgap determining charge dynamics. Extended defects, including grain boundaries and material interfaces, contribute significantly by hosting high densities of charge-trapping states that influence leakage currents and nonlinear behavior. Material properties such as microstructures, band gap energy profiles, and chemical composition dictate the density, distribution, and activity of these defects, shaping the overall dielectric response. Accurate modeling requires the incorporation of defect density, charge trapping/detrapping kinetics, and space charge effects, alongside interface phenomena, to comprehensively capture the interplay between defects and polarization.

\begin{figure}
    \centering
    \includegraphics{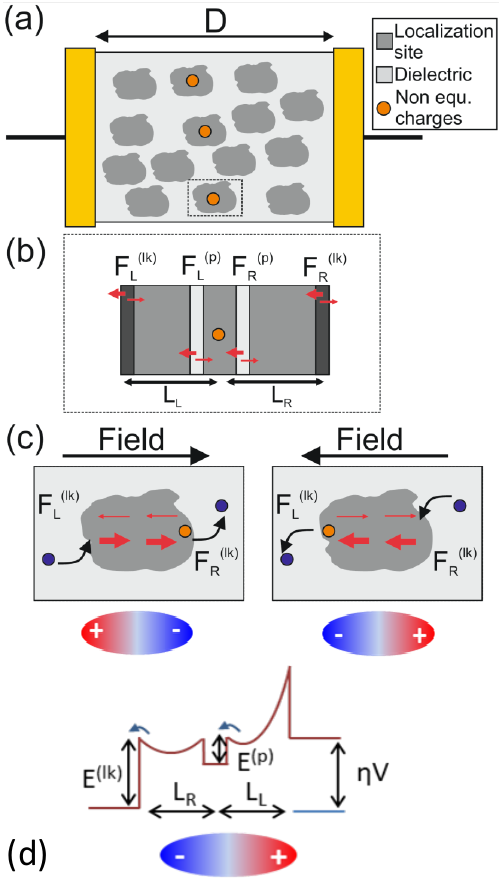}
    \caption{(a) Illustration of a capapacitor with plate separation $D$ and plate area $A$ with a dielectric medium containing charge localization sites of mean size $L$. (b) Barrier profile showing the carrier fluxes overcoming $E^{(p)}$ barriers that lead to the charge bouncing within the site and the carrier fluxes overcoming $E^{(lk)}$ barriers that lead to the charge leak. (c) Illustration of internal fluxes of nonequilibrium charges leading to polarization fluctuations with the presence of leaking processes as fluxes overcoming the external barriers into the sorrounding dielectric. (d) Diagram summarizing the polarization transfer within confinement sites and the charge leakage pathways, with the associated potential barriers $E^{(p)}$ and $E^{(lk)}$, along with the role of symmetry.}
    \label{fig:01}
\end{figure}

To analyze the effects just highlighted, we developed a model consisting of a capacitor filled with a dielectric material of size $D$, as represented in Fig.~\ref{fig:01} (a). Inside this material, charges can be trapped in small sites of mean size 
$L$, which are displaced from equilibrium when subjected to an external electric field. These sites can have different shapes and size distributions that could correspond to crystallites in a polycrystalline material or grains in an amorphous one. 
Different dynamics of charges within the capacitor are considered in this
model:
\begin{itemize}
\item  Charges can enter the system and be trapped in
any site.
\item  Charges can escape from a site due to a leakage
effect and become trapped by another site.
\item  Charges can escape from a site due to a leakage
effect and exit the system.
\item  A small amount of charges can still enter and exit
the system directly without undergoing any of the
previous steps.
\end{itemize}

Under an electric field, as illustrated in Fig.~\ref{fig:01} (b) fluxes of nonequilibrium charges can produce a local polarization imbalance as mobile charges bounce within the sites following the voltage polarity. Additionally, some amount of charges can leak out from the bouncing sites, as represented in  Fig.~\ref{fig:01} (c).

The nonequilibrium charge fluctuations can be produced by localized charge traps or sources (such as a point defect or array of defects) present within the localization sites of size $L$, as they are biased at a fraction $\eta V$ of the total applied voltage $V$. In the context of a uniform distribution of sites, $\eta =L/D$ represents the inversed number of sites along a line between the two contacts. Thus, the parameter $\eta$ characterizes the voltage drop efficiency per site.

Out of equilibrium, the polarization transfer function can be derived from the sum of time rates of local mobile charge generation or trapping, $\partial q_i/\partial t $, at a position  $\langle r_i \rangle$ over the total volume, $\Omega=D \cdot A$, as follows
\begin{equation}
    \left. \frac{\partial P}{\partial t} \right|^{(\gamma)}=\frac{1}{\Omega} \sum_{i} \frac{\partial q_i}{\partial t} \langle r_i \rangle,
    \label{eq0}
\end{equation}
that leads to electric dipole, $\partial q_i\langle r_i \rangle$, generation through polarization confinement ($\gamma=p$) or dipole annihilation through leaking processes ($\gamma=lk$). Note that in this framework, we are assuming fixed dipole orientations and neglecting any geometric reshaping, such as rotation. By assuming a uniform distribution of sites within an area $A$ the sum in Eq.~\ref{eq0} can be segmented over the inplane dipoles, $k$, and along the distance perpendicular to the capacitor plates, $j$, as
\begin{equation}
    \left. \frac{\partial P}{\partial t} \right|^{(\gamma)}=\frac{1}{D} \sum_{j} \frac{1}{A} \sum_k \frac{\partial q_j^{(k)}}{\partial t} \langle r_j \rangle,
    \label{eq1}
\end{equation}
leading to a dependence of the polarization transfer on a combination of local charge fluxes $F^{(\gamma)} \equiv 1/A \sum_k \partial q_j^{(k)}/\partial t$.

We assume that the charge within each site redistributes either to the left or right according to the local bias, which is proportional to the total applied voltage $\eta V$, following the configuration shown in Fig.~\ref{fig:01}(b). Thus, the nonequilibrium polarization transfer described in Eq.~\ref{eq1} will induce a local bouncing of carriers, $\partial P/\partial t |^{(p)}$, which can be characterized in terms of carrier fluxes over left and right internal barriers. Concurrently, charges can either leak or enter the site via the external barriers, leading to a polarization fluctuation rate $\partial P/\partial t |^{(lk)}$. In a first approximation, we will assume these leaking mechanisms to be independent of the bouncing process, each with potentially distinct relaxation times, $\tau^{(p)}$ and $\tau^{(lk)}$.

Given the total number of localization sites along a line between contacts, proportional to $1/\eta$, for carriers with unit charge $q$ (where the sign may vary depending on the nature of the charge: electrons, holes, ions), the expression in Eq.~\ref{eq1} can be simplified to
\begin{equation}
    \left. \frac{\partial P}{\partial t} \right|^{(\gamma)}=\frac{1}{D \eta}  q \left(F_L^{(\gamma)} L_L + F_R^{(\gamma)} L_R \right),
    \label{eq2}
\end{equation}
where $F_L^{(\gamma)}$ and $F_R^{(\gamma)}$, represent the carrier fluxes to the left and right-hand side, respectively, as shown in Fig.~\ref{fig:01} (a). And charges may accumulate (or leak) at a distance $L_R$, to the right, or $L_L$, to the left, of the generation center. This configuration also accounts for the possibility of an asymmetric energy difference between the localized charge source and the left and right sides of the site. This asymmetry can be represented by a symmetry parameter $\alpha \in [0,\infty]$, such that $\eta_L=\alpha \eta/(1+\alpha)$ and $\eta_R=\eta/(1+\alpha)$.  This asymmetry can also be spatially projected, with $L_L=\alpha L/(1+\alpha)$ and $L_R=L/(1+\alpha)$, where $\alpha=1$ corresponds to the perfectly symmetric case.

\begin{figure}
    \centering
    \includegraphics{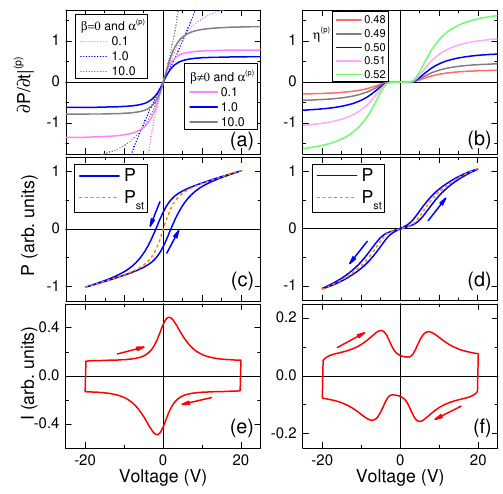}
    \caption{(a) Contrasting transfer functions as function of applied bias for $\alpha<1$, $\alpha=1$, and $\alpha>1$ with (without) the saturation factor $\beta \neq 0$ ($\beta=0$), for $\eta=0.005$ and period $T=10 \pi \tau^{(p)}$. (b) Contrasting nonequilibrium charge transfer functions as function of applied bias for varying efficiency factors $\eta$, $\alpha=1.0$, and period $T=20 \pi \tau^{(p)}$. (c) Calculated polarization ($P$) as a function of the applied bias for triangular voltage sweeps, of large voltage amplitudes, in the absence of charge leakage, under parameters of the transfer function (a) and (d) calculated polarization ($P$) under parameters of the transfer function (b). The stationary results ($P_{st}$) for the polarization-voltage characteristics have been represented using red dashed curves. The respective currents (I) are provided in panels (e) and (f).}
    \label{fig:02}
\end{figure}

In the case of electrons sources with $q=-e$ and effective mass $m^*$ at an effective temperature $T_{eff}$~\cite{Guarin2021}, the corresponding fluxes due to thermal ionization are given by
\begin{equation}
    F_{L(R)}^{(\gamma)}=\lambda^{(\gamma)} \left[1- \exp\left( \eta_{L(R)} \frac{eV}{k_B T_{eff}} \right)  \right],
    \label{eq3}
\end{equation}
and
\begin{equation}
   F_{R(L)}^{(\gamma)}=\lambda^{(\gamma)} \left[\exp\left( -\eta_{R(L)} \frac{eV}{k_B T_{eff}} \right) -1 \right],
    \label{eq4}
\end{equation}
with $\lambda^{(\gamma)} = 4 \pi m^* (k_B T_{eff})^2 \exp\left( -\frac{E^{(\gamma)}}{k_B T_{eff}} \right)/(2 \pi \hbar)^3$. The kinetics of charge trapping and detrapping at each localization site are modeled using thermally activated Maxwell-Boltzmann statistics, which describe the flux of particles across energy barriers. Thus, the nonequilibrium fluxes are tuned by internal activation barriers, $E^{(p)}$, attributed to the carrier source or trapping that leads to the charge bouncing within the site. Simultaneously, an external barrier, $E^{(lk)}$, is also present and must be overcome for charge to leak out or be trapped into the site. These processes are illustrated in Fig. \ref{fig:01} (d).

The eventual exhaustion of the trapping or generation process, which drives the polarization fluctuation at large voltages (independent of its polarity), can be modeled by saturating the local voltage efficiency, just for the polarization transfer and not the charge leakage. This is achieved by substituting $\eta \rightarrow \eta/\sqrt{1+\beta V^2}$, in Eqs.~\ref{eq3} and~\ref{eq4} for $(\gamma=p)$ introducing $\beta > 0$ as an adjustable parameter so that this effect does not contribute at small amplitudes when $\beta V^2<< 1$. 

The polarization transfer functions, derived by combining Eqs.~\ref{eq2}-\ref{eq4}, are presented in Fig.~\ref{fig:02} (a) for various values of the symmetry parameter $\alpha$. These functions consistently exhibit a monotonic increase, where the correlation between voltage polarity and their sign remains fixed: positive polarity corresponds to a positive sign, while negative polarity results in a negative sign. The breaking of symmetry, as reflected by varying $\alpha$, leads to an asymmetry in the function’s strength with respect to polarity. Additionally, the impact of polarization transfer exhaustion, either included ($\beta \neq 0$) or excluded ($\beta = 0$), is also displayed in panel~\ref{fig:02} (a), highlighting its influence on the overall behavior of the transfer function.

Moreover, increasing the voltage drop efficiency $\eta$ causes the apparent onset of the transfer function to extend beyond $V=0$, as depicted in Fig.~\ref{fig:02} (b). Next, we will examine how these effects influence the tuning of electric polarization dynamics and the corresponding current-voltage characteristics.

The presence of polarization transfer, described by Eq.~\ref{eq2}, will result in fluctuations of the polarization around certain instantaneous equilibrium values determined by the instantaneous electric susceptibility, $\chi_{\infty}$. The instantaneous polarization is given by $P_{\infty}=\varepsilon_0 \chi_{\infty} E$, $\varepsilon_0$ being the vacuum permittivity, assuming a uniform electric field $E= V/D$ along the device, so the total polarization can be expressed as 
\begin{equation}
P=P_{\infty}+\delta P^{(p)} - \delta P^{(lk)}. 
\label{eq5}
\end{equation}
Within the relaxation time, $\tau^{(\gamma)}$, approximation, these fluctuations can be described as
\begin{equation}
   \frac{d \delta P^{(\gamma)}}{d t}=-\frac{\delta P^{(\gamma)}}{\tau^{(\gamma)}}+\left. \frac{\partial P}{\partial t} \right|^{(\gamma)}.
    \label{eq6}
\end{equation}
With these components, the displacement current can be calculated as
\begin{equation}
    j_D = \varepsilon_0 \frac{dE}{dt} + \frac{dP}{dt}.
    \label{eq7}
\end{equation}

To calculate the total current through the device, we also account for a residual carrier drift with conductance $G_0$ complemented by the leaking charge fluctuation $\delta n$
\begin{eqnarray}
    \label{eq8}
    I &=& j_{D} \cdot A + \left( G_{0} + \Gamma \delta n \right) V \\
      &=& C \frac{d V}{d t}+\frac{d \delta P^{(p)}}{dt} - \frac{d \delta P^{(lk)}}{dt}  + \left( G_{0} + \Gamma \delta n \right) V. \nonumber
\end{eqnarray}

\begin{figure}
    \centering
    \includegraphics{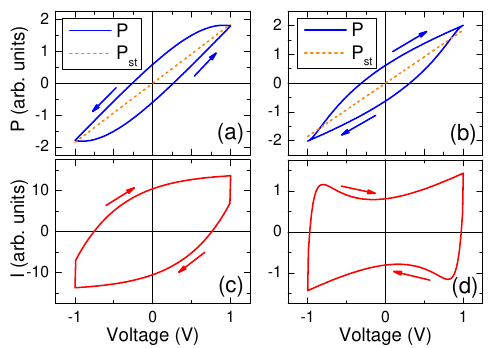}
    \caption{Small voltage amplitude polarization ($P$) in presence of both, polarization fluctuation and leakage with $\tau^{(lk)}/\tau^{(p)}=10$ with the parameters listed in Table~\ref{table} of the Appendix: (a) for a voltage sweep period, $T=2 \pi \tau^{(p)}$; (b) for a voltage sweep period, $T=2 \pi \tau^{(lk)}$. The stationary results ($P_{st}$) for the polarization-voltage characteristics have been represented using red dashed curves. Their respective currents (I) are displayed in panels (c) and (d).}
    \label{fig:03}
\end{figure}

Here, $C = \varepsilon_{0}(1+\chi_{\infty}) A/D$, $\Gamma = e \mu/ D^2$, and $\mu$ represents the nonequilibrium carrier mobility~\cite{Paiva2022}. The potential effects of drift current modulation with nonequilibrium charge fluctuations, $\delta n$, are detailed in Refs.~\cite{LopezRichard2022,lopez2024beyond, lopez2024accuracy, Lopez-Richard2024}. Therefore, for the purposes of this study, we will assume $ \Gamma \delta n<< G_{0}$, maintaining just a residual leaking conductance, $G_{0} \neq 0$, in addition to the displacement contributions. For simplicity, we have set $G_0=1$ as the unit of conductance in the presence of leakage.

Figure \ref{fig:02} (c) illustrates an example of hysteresis in the polarization-voltage loop under a periodic triangular input, calculated according to Eq.\ref{eq5} for a symmetrical transfer function with $\alpha=1$ and in the absence of leakage. The arrows indicate the direction of the loop. Notably, the generated polarization resembles a ferroelectric response~\cite{kumar2021induced,Sekine2022,Manzi2023}. The polarization is not fully saturated due to the linear voltage dependence assumed for $P_{\infty}=\varepsilon_0 \chi_{\infty} V/D$. When increasing the parameter $\eta$, under a symmetric transfer function, the voltage onset that triggers polarization fluctuation increases, resulting in an apparent pinched polarization hysteresis, as shown in panel~\ref{fig:02} (d). This resembles the traces of a double hysteresis loop, which are typically associated with antiferroelectric behavior~\cite{pintilie2008coexistence,wang2020paraelectric,luo2023well}. In this scenario, the apparent double hysteresis loops emerge due to the widening voltage onset of the transfer function. The stationary result for the polarization-voltage characteristics, $P_{st}$, obtained by setting $d \delta P^{(p)}/dt=0$ in Eq.~\ref{eq5} has been added to both Figs~\ref{fig:02} (c) and (d) as reference by using dashed curves.

The corresponding current-voltage loops, calculated according to Eq.~\ref{eq8}, are depicted in panels~\ref{fig:02} (e) and (f), respectively. The opening of the hysteresis response depends not only on the relative magnitude of nonequilibrium contributions ($\delta P$ in this case) compared to the stationary values, but also on the period of the cyclic voltage. Optimal conditions for a memory response, when the voltage period approaches $T \sim 2 \pi \tau^{(\gamma)}$, are discussed in Refs.~\citenum{Paiva2022,Silva2022,LopezRichard2022}.   

For small voltage amplitudes, the saturation effects can be neglected ($\beta V^2\approx 0$, after Eqs.~\ref{eq3} and~\ref{eq4}) and solving Eqs.~\ref{eq2}-\ref{eq8} in the presence of leakage, one may obtain results as those depicted in Figs~\ref{fig:03} (a)-(d). In order to contrast the potential time-scale difference we have set $\tau^{(p)}=\tau^{(lk)}/10$ with symmetric polarization transfer ($\alpha=1$).
In Fig.~\ref{fig:03} (a), a ferroelectric-like counterclockwise loop is achieved by setting the voltage sweeping period to $T = 2 \pi \tau^{(p)}$, which maximizes polarization fluctuations while keeping the leakage contribution negligible. By adjusting the voltage sweep period to $T = 2 \pi \tau^{(lk)}$, the polarization loop direction can reverse, as shown in Fig.~\ref{fig:03} (b). Under this condition, the voltage sweep aligns with the charge leakage dynamics, enabling the hysteresis loop to switch direction. While the (p) component alone is sufficient to produce ferroelectric-like loops by enhancing net polarization through space charge accumulation, the inclusion of the (lk) component is critical to explain phenomena such as polarization loop inversion, thereby providing a more comprehensive understanding of the interplay between charge trapping and leakage processes in leaky dielectrics. The phenomenon of hysteresis loop inversion arises from the interplay between charge trapping, leakage currents, and the time-dependent asymmetries in polarization dynamics. Intuitively, when the leakage current dominates over the trapped charge contribution, it effectively reverses the net polarization direction during the cycle. This behavior can be visualized as the system's inability to maintain the accumulated polarization under certain conditions, leading to an apparent inversion of the loop. Such inversion is driven by competing processes: while trapped charges enhance polarization, leakage acts to dissipate it. The timing and magnitude of these processes relative to the applied field result in the observed inversion. This intricate balance highlights the critical role of system parameters such as trap density, field frequency, and material anisotropy, which collectively govern the polarization dynamics and shape the hysteresis response. Experimentally, clockwise reversible polarization loops have been observed in Ref.~\citenum{Wei2005}, and similar effects have been investigated in Ref.~\citenum{Jin2023}. 

\begin{figure}
    \centering
    \includegraphics[scale=1]{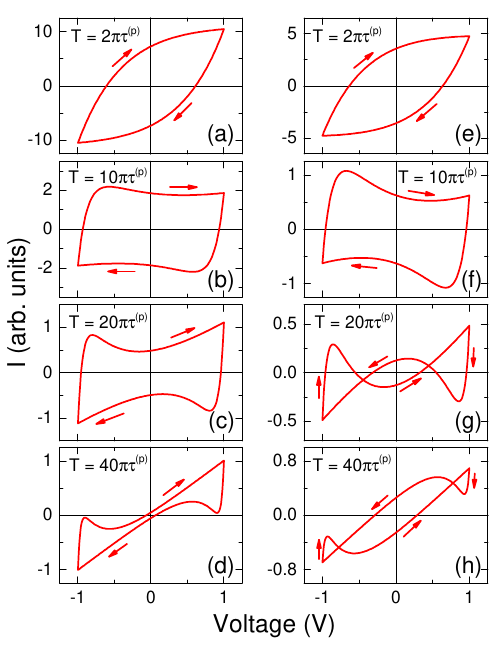}
    \caption{Left column: current-voltage loops using the same parameters as in Fig.~\ref{fig:03} (a) as listed in Table~\ref{table} of the Appendix, but with capacitance $C\rightarrow 0$, varying the period, $T$. Right column: current-voltage loops with the same parameters but $\lambda^{(p)}=50.0$, varying the period, $T$.}
    \label{fig:current}
\end{figure}

Reference \citenum{Wei2005} investigated barium stannate titanate ceramics (BaTi$_{0.9}$Sn$_{0.1}$O$_3$, BTS) using capacitance-voltage ($C$-$V$) and Sawyer-Tower circuit measurements, revealing abnormal clockwise hysteresis loops in the temperature range of 10–40 °C. These loops were attributed to the strong interaction between point defects and domain walls, which limited domain switching and induced reversible polarization dynamics. Similarly, Jin et al. \citenum{Jin2023} explored ferroelectric field-effect transistors (FeFETs) based on Hf$_{0.5}$Z$r_{0.5}$O$_2$, focusing on polarization switching and interface trap effects. In this system, clockwise hysteresis loops were linked to trapped charges at the channel/ferroelectric interface, which opposed polarization-induced voltage shifts.

Both studies underscore the critical role of defects and interfacial phenomena in driving hysteresis loop inversion. Our model reproduces these behaviors by incorporating trapping/detrapping kinetics, space charge effects, and the interplay between polarization transfer and leakage currents. For instance, the temperature-dependent response of BTS \cite{Wei2005} and the interface-driven dynamics in FeFETs \cite{Jin2023} align well with our framework. These experimental setups also highlight the significance of material-specific properties, including defect densities, band alignment, and microstructural factors, in tuning dielectric responses. By bridging microscopic mechanisms and macroscopic observations, our model complements these findings and provides a predictive tool for designing and optimizing materials with tailored dielectric and ferroelectric-like behaviors.

The term \textit{proteresis} has been coined~\cite{Girard1989} to describe these kind of loops characterized by an apparent anticipation of the response~\cite{Wei2008,Syed2020,Khan2021}, in contrast to hysteresis (coming from the Greek meaning something that comes later~\cite{Girard1989}), which would represent delayed dynamics. While this terminology aims to capture the opposite nature of such loops, it can lead to the misconception that the apparent anticipation suggests a lack of causality. Our model demonstrates that the direction of the loop is not related to the perturbation timing but it is fully attributable to the nature of the transfer function in Eq.~\ref{eq2}, whether it involves carrier trapping or release, both of which are inherently delayed processes governed by specific relaxation times according to Eq.~\ref{eq6}. For this reason, we advocate that using the term inverted hysteresis remains accurate and appropriate, contrary to suggestions made elsewhere.

In addition to frequency tuning, the balance between polarization transfer and charge leakage is significantly influenced by temperature, governed by the parameters $\lambda^{(p)}$ and $\lambda^{(lk)}$ as defined after Eq.~\ref{eq4}, as confirmed in Ref.~\citenum{Wei2005}. The impact of temperature tuning on these effects will be further addressed in the discussion.

Note that despite the reversal of the polarization loop, the current loops have consistently maintained a clockwise direction throughout both Fig.~\ref{fig:02} and Fig.~\ref{fig:03}. This behavior arises from the relatively strong contribution of the capacitive term in Eq.~\ref{eq8}, $C dV/dt$. For a negligible capacitive contribution (with $C \rightarrow 0$), a reversal of the current loops is also expected. In Fig.~\ref{fig:current}, we compare two scenarios with parameters minimizing the capacitive term: the left column shows a larger value of $\lambda^{(p)}$, where no loop reversal is observed within the analyzed period range, while for a smaller $\lambda^{(p)}$, a counterclockwise current loop emerges as the period increases. This effect has been experimentally observed in Ref.~\citenum{Jin2023} and attributed to the detrapping of electrons from interface traps, in agreement with the theoretical mechanisms described in our model.

%%%%%% nova discusão aqui
Given the period tunability of the polarization and displacement response, a spectroscopic analysis becomes paramount for a comprehensive characterization of the transport behavior. By analyzing the frequency-dependent response, we can better understand how charge dynamics, polarization transfer, and leakage processes interact across different time scales. This allows us to probe the underlying mechanisms in greater detail and assess the material's overall dielectric and transport properties more effectively.

Under a sinusoidal voltage input, $V=V_0 \cos(\omega t)$, and by expanding Eqs.~\ref{eq2}-\ref{eq4} in powers of $\eta eV_0/(k_B T_{eff})$ the solution of Eq.~\ref{eq6} can be derived analytically.
The stable response, once all transient contributions have dissipated, can be expressed as a combination of harmonics \cite{lopez2024beyond}
\begin{equation}\label{eq9}
    I = V_{0}\sum_{n=0}^{}\left[G_{\left(n\right)} \cos\left(n\omega t\right) - B_{\left(n\right)}\sin\left(n\omega t\right)\right].
\end{equation}
The higher harmonic response originates from the nonlinear functional dependence of the generation function on the applied bias. While the first-order response, including the imaginary reactive part, produces elliptical Lissajous loops in polarization-voltage characteristics, any deviation from these simple elliptical loops signifies nonlinear and thus higher-order contributions. Thus, assuming $\eta eV_0/(k_B T_{eff})<<1$ the $n-mode$ conductance $G_{(n)}$ and susceptance $B_{(n)}$, up to second order, are given by
\begin{equation}\label{eq10}
    G_{(1)} = \frac{A}{V_0}\left\{\frac{\sigma_{1}^{(p)}\left[\omega \tau^{(p)}\right]^2}{1+\left[\omega \tau^{(p)}\right]^{2}} - \frac{\sigma_{1}^{(lk)}\left[\omega \tau^{(lk)}\right]^2}{1+\left[\omega \tau^{(lk)}\right]^2}\right\} + G_0,
\end{equation}

\begin{figure}
    \centering
    \includegraphics[scale=1]{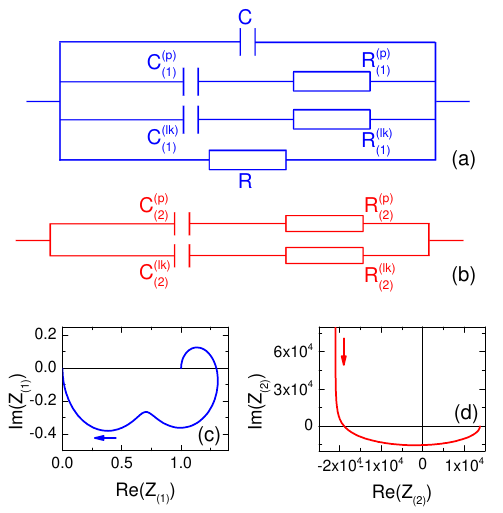}
    \caption{Apparent circuit representations for: (a) the first and (b) the second mode of the dielectric impedance. First (c) and second (d) order mode Nyquist plots for: $G_0 = 1.0$ and $C = 0.01$. The rest of the parameters is listed in Table~\ref{table} of the Appendix. The arrows point towards the frequency growth direction.}
    \label{fig:04}
\end{figure}

\begin{equation}\label{eq11}
    B_{(1)} = \frac{A}{V_0}\left\{\frac{\sigma_{1}^{(p)}\omega \tau^{(p)}}{1+\left[\omega \tau^{(p)}\right]^2} - \frac{\sigma_{1}^{(lk)}\omega \tau^{(lk)}}{1+\left[\omega \tau^{(lk)}\right]^2}\right\} + \omega C,
\end{equation}
and
\begin{equation}\label{eq12}
    G_{(2)} = \frac{2A}{V_0}\left\{\frac{\sigma_{2}^{(p)}\left[\omega \tau^{(p)}\right]^2}{1+\left[2\omega \tau^{(p)}\right]^2} - \frac{\sigma_{2}^{(lk)}\left[\omega \tau^{(lk)}\right]^2}{1+\left[2\omega \tau^{(lk)}\right]^2}\right\} ,
\end{equation}
\begin{equation}\label{eq13}
    B_{(2)} = \frac{A}{V_0}\left\{\frac{\sigma_{2}^{(p)}\omega \tau^{(p)}}{1+\left[2\omega \tau^{(p)}\right]^2} - \frac{\sigma_{2}^{(lk)}\omega \tau^{(lk)}}{1+\left[2\omega \tau^{(lk)}\right]^2}\right\} ,
\end{equation}
where
\begin{equation}\label{eq14}
    \sigma_{1}^{(\gamma)} = \lambda^{(\gamma)} \frac{\left[\alpha^2 + 1\right]}{\left[1+\alpha\right]^2} \frac{\eta eV_0}{k_B T_{eff}},
\end{equation}
and
\begin{equation}\label{eq15}
    \sigma_{2}^{(\gamma)} = \frac{\lambda^{(\gamma)}}{2} \frac{\left[\alpha^{3} - 1\right]}{\left[1+\alpha\right]^3} \left[\frac{\eta eV_0}{k_B T_{eff}}\right]^2.
\end{equation}
Note that, for $n=1$ the charge bouncing contribution, proportional to $\sigma^{(p)}_1$, is positive defined while the leakage term, proportional to $\sigma^{(lk)}_1$, is negative. This will be the cause of the apparent inductive traces in the fundamental mode if $B_{(1)}<0$.   

In this representation, the impedance per mode can be expressed as~\cite{lopez2024beyond,lopez2024accuracy}
\begin{equation}\label{impedance}
    Z_{(n)} = \frac{G_{(n)}}{[G_{(n)}]^{2} + [B_{(n)}]^{2}} - i\frac{B_{(n)}}{[G_{(n)}]^{2} + [B_{(n)}]^{2}}.
\end{equation}
This formulation provides an impedance spectroscopic characterization that extends beyond the traditional fundamental mode,~\cite{wang2021electrochemical, lazanas2023electrochemical} offering deeper insights into the system's behavior. Multimode impedance spectroscopy, by analyzing higher harmonics in the current response, provides access to information beyond that obtainable from conventional linear impedance spectroscopy ~\cite{lopez2024beyond,lopez2024accuracy}. This technique allows for the characterization of non-linear phenomena and can be particularly valuable for understanding the behavior of complex systems, such as those found in energy conversion devices \cite{Lopez-Richard2024}, memory devices \cite{lopez2024beyond}, and advanced materials \cite{lopez2024emergence,ames2024optical}.

According to the frequency dependence described by Eqs.~\ref{eq10}- \ref{eq13}, the apparent circuits for the first and second mode impedance is depicted in Figs.~\ref{fig:04} (a) and (b), respectively. Note that, for a resistance, $R'$, in series with a capacitor, $C'$ the effective admittance is
\begin{equation}
    Y = \frac{1}{R'}\left[\frac{\left(\omega \tau' \right)^{2}}{1+\left(\omega \tau' \right)^{2}}+i\frac{\omega \tau'}{1+\left(\omega \tau'\right)^{2}}  \right],
\end{equation}
with $\tau' =R' C'$. Thus, following Eqs.~\ref{eq10} and \ref{eq11}, the apparent resistances and capacitances in panel~\ref{fig:04} (a) are defined as follows:  $R\equiv 1/G_0$, $R^{(p)}_1=V_0/(A \sigma^{(p)}_1)$, and $C^{(p)}_1=\tau^{(p)}/R^{(p)}_1$, while $R^{(lk)}_1=-V_0/(A \sigma^{(lk)}_1)$, and $C^{(lk)}_1=-\tau^{(lk)}/R^{(lk)}_1$. Note that the admittance results, when contrasted with apparent circuit representations, may lead to parameters such as negative capacitances, unexpected inductances, or unusual frequency dependencies, as observed in a vast body of published studies \cite{Priyadarshani2021,Bou2020,Abdulrahim2021,Gonzales2022,Thapa2022,Joshi2020}. As discussed in Ref. \cite{lopez2024beyond}, these are metaphorical representations of potentially more complex responses. Hence, establishing a microscopic correlation between these circuit elements and actual system parameters is essential for accurate interpretation. In turn, for the second order term, the apparent circuit is represented in panel~\ref{fig:04} (b). The resistances and capacitances in this case, according to Eqs.~\ref{eq12} and \ref{eq13}, are defined as $R^{(p)}_2=2V_0/(A \sigma^{(p)}_2)$, and $C^{(p)}_2=2 \tau^{(p)}/R^{(p)}_2$, while $R^{(lk)}_1=-2 V_0/(A \sigma^{(lk)}_2)$, and $C^{(lk)}_2 =-2\tau^{(lk)}/R^{(lk)}_2$. In this case, the ambiguity in the values of $\alpha \neq 1$ in the presence of asymmetry prevents a strict sign definition for $\sigma^{(\gamma)}_2$ in Eq.~\ref{eq15} and, consequently, for the sign of the apparent resistance or capacitance of the second mode branch. 

Two examples of multimode impedance maps for the first and second order modes have been displayed in Figs.~\ref{fig:04} (c) and (d), respectively. Finite values of the residual conductance $G_0$ and the geometric capacitance, $C$, have been considered assuming $\tau^{(lk)}=10 \tau^{(p)}$. 
For $n=1$, at low frequencies, an apparent inductive loop emerges, as shown by the positive values of $\text{Im}(Z_{(1)})$ in panel~\ref{fig:04}(c). These effects are often referred to as negative capacitance \cite{cheng2019negative}, as they account for seemingly inductive traces in the absence of real inductance. In this case, they emerge from the interplay between opposing contributions within the system, reflecting the nonlinear and memory effects embedded in the microscopic dynamics. Specifically, the opposing signs of the charge storage (bouncing) and leakage terms in Eq.~\ref{eq11} lead to this behavior. This dynamic gives rise to inductive-like responses without invoking actual inductance, manifesting as apparent negative capacitance in the impedance spectrum.

\begin{figure}
    \centering
    \includegraphics[scale=1]{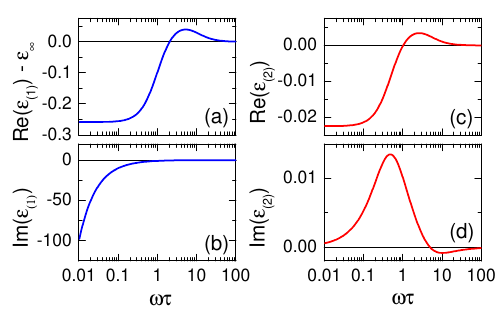}
    \caption{First mode contributions to the real (a) and imaginary (b) part, of the dielectric permittivity, and the corresponding second-order contributions to the real (c) and (d), and imaginary parts for the set of parameters used in Fig.~\ref{fig:04} (c) and (d) as listed in Table~\ref{table} of the Appendix. }
    \label{fig:05}
\end{figure}

The contribution of the second mode, shown in panel ~\ref{fig:04}(d), exhibits a noticeable transition in the impedance character as the frequency increases, shifting from the second to the fourth quadrant. This behavior arises as both the real and imaginary components of the impedance reverse sign. The transition is driven by an evolving imbalance between the opposing contributions of charge leakage and charge bouncing. As frequency increases, the dominance of one mechanism over the other shifts, leading to this distinctive impedance response. This highlights the complex interplay between dynamic processes that cannot be captured by simpler, first-order approximations.

In analyzing the transport response of a dielectric, it is important to note the correlation between impedance spectroscopy and dielectric spectroscopy.  While impedance spectroscopy provides insights into the resistive (real part) and reactive (imaginary part) components of impedance, dielectric spectroscopy measures the dielectric permittivity and loss as a function of frequency, offering detailed information about dipole relaxation processes and polarization mechanisms within the material. In our model, these techniques are connected through the multimode expression for the electric susceptibility in nonlinear systems

\begin{equation}
    P=P_{\infty}+\varepsilon_0 \sum_{n=1}\chi_{(n)}E^n.
\end{equation}
Thus, by linking this expansion with Eqs.~\ref{eq8} and~\ref{eq9}, we can generalized the permittivity definition in this multimode perspective (including the residual conductance) as
\begin{equation}
   \varepsilon_{(n)}=\varepsilon_0 \chi_{(n)} +\delta_{n,1}\left[\varepsilon_0(1+\chi_{\infty})  -i \frac{G_0}{\omega} \frac{D}{A}\right].
   \label{epsi}
\end{equation}
The current in Eq,~\ref{eq8} can be thus redefined in the complex plane as
\begin{equation}
   I=A \sum_{n=1} i \varepsilon_{(n)}  n \omega \left( \frac{V_0}{D}\right)^n e^{i n \omega t},
\end{equation}
that, according to Eq.~\ref{eq9} leads to the following correlation between the permittivity, the conductance, and susceptance per mode 
\begin{equation}
    \varepsilon_{(n)} = \frac{[B_{(n)} - iG_{(n)}]}{n\omega} \frac{D^n}{A V_{0}^{n-1}},
    \label{epsin}
\end{equation}
for $n=1,2,...$
The real and imaginary contributions to the first and second-order terms of the electric permittivity, corresponding to the parameters in Figs.~\ref{fig:04} (c) and (d), are depicted in Fig.\ref{fig:05}. The behavior expected for the real part of the $n=1$ permittivity is $\text{Re} [\varepsilon_{(1)}] \propto B_{(1)}/\omega$ in Eq.~\ref{epsin}. This results in step-up or step-down functions as a function of frequency (in logarithmic scale) depending on the leak or polarization contributions, respectively, as shown in Fig.~\ref{fig:05} (a). Due to the presence of residual conductance, the corresponding imaginary part in panel \ref{fig:05} (b) diverges for decreasing $\omega$ since $\text{Im} [\varepsilon_{(1)}] \propto -G_0/\omega$ in Eq.~\ref{epsi}. For the second mode, step functions (in logarithmic scale) are also expected for $\text{Re} [\varepsilon_{(2)}] \propto B_{(2)}/\omega$ for the same reasons as for the first mode, as depicted in Fig.~\ref{fig:05} (c). The imaginary parts in this case collapse at low and high frequencies and peak at $\omega = 1/(2 \tau^{(lk)})$.

Higher harmonics in the dielectric response offer crucial insights into a range of phenomena with direct implications for device performance \cite{lopez2024beyond,lopez2024accuracy}. In ferroelectric materials, they uncover domain wall dynamics, which are critical for optimizing memory applications \cite{jimenez2020modeling}. In multiferroic systems, higher harmonics probe magnetoelectric coupling, enhancing the understanding of multifunctional devices \cite{zemp2024magnetoelectric}. For energy storage applications, they elucidate electric and dielectric relaxation mechanisms, providing pathways for improving efficiency and stability \cite{Lopez-Richard2024}. Additionally, higher harmonics play a pivotal role in studying protein dynamics in biosensors \cite{matyushov2023nonlinear} and in characterizing novel metamaterials with unique electromagnetic properties \cite{wang2022wireless,marquardt2023domain}. These examples demonstrate the broad applicability of higher harmonic analysis, highlighting its potential to optimize existing technologies and drive the development of next-generation materials and devices.

As discussed earlier, the relative contributions of polarization transfer and charge leakage (captured by the parameters $\lambda^{(\gamma)}$ after Eq.~\ref{eq4}) can be modulated by varying the temperature. This temperature-dependent tuning of the dielectric response, combined with frequency adjustments, offers valuable insights into the dynamic behavior of the dielectric material. To analyze this further, we can utilize the explicit expression for the first-mode permittivity according to Eqs.~\ref{epsin} and~\ref{eq11} 
\begin{widetext}

\begin{equation}
\text{Re} \left[ \varepsilon_{(1)} \right] = \varepsilon_0(1+\chi_{\infty}) 
+ \frac{4 \pi m^* D e}{(2 \pi \hbar)^3}  k_B T_{eff} \frac{(1+\alpha^{2})}{(1+\alpha)^2}\eta \left[  \frac{ \tau^{(p)}}{1+\left[\omega \tau^{(p)}\right]^2} e^{-\frac{E^{(p)}}{k_B T_{eff}}} 
-  \frac{ \tau^{(lk)}}{1+\left[\omega \tau^{(lk)}\right]^2} e^{-\frac{E^{(lk)}}{k_B T_{eff}}} \right] 
\label{perm}
\end{equation}

\end{widetext}

\begin{figure}
    \centering
    \includegraphics{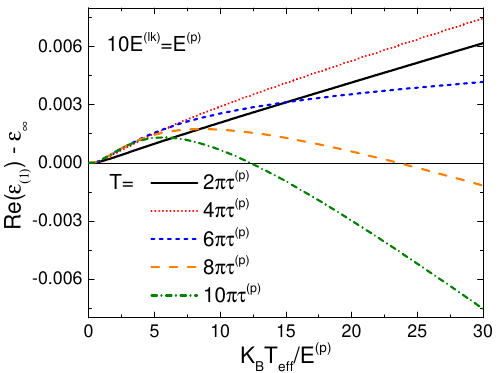}
    \caption{First mode dielectric permittivity as a function of temperature, $T_{eff}$ for increasing AC voltage period and the parameters listed in Table~\ref{table} of the Appendix.}
    \label{fig:06}
\end{figure}
that reveals a peculiar temperature tuning behavior in presence of leakage. By establishing contrasting relaxation times for polarization fluctuation and charge leakage, $\tau^{(p)}< \tau^{(lk)}$, along with different values for the potential barriers, $E^{(p)}<E^{(lk)}$, the temperature dependence of this function can shift. As displayed in Fig.~\ref{fig:06}, at high frequencies $\omega$ (low period $T=2 \pi/\omega$) it exhibits a monotonic trend, but as the frequency decreases, it transforms into a non-monotonic pattern resembling a relaxor response~\cite{Ramesh2016,Zhou2024}. At low temperatures, the dielectric tuning of polarization transfer dominates, while at higher temperatures, the leakage contribution, which reduces the relative permittivity as described in Eq.~\ref{perm}, becomes more significant.

To experimentally validate the theoretical predictions presented in this study, several approaches can be employed. Firstly, temperature-dependent polarization-voltage ($P$-$V$) measurements can be conducted to investigate the role of thermal activation in polarization dynamics. The observed temperature dependence of the hysteresis loop shape, including potential inversions under specific conditions, and size, due to the peculiar temperature scaling, will provide crucial evidence supporting the model's predictions.

Secondly, frequency-dependent impedance spectroscopy combined with dielectric spectroscopy will offer valuable insights. Transitions in the real part of the permittivity can be correlated with internal time scales associated with defect-related processes and ferroelectric relaxation. Furthermore, analyzing higher harmonic responses will enable the identification of signatures of non-equilibrium charge trapping and leakage processes, such as the presence of apparent negative capacitance or inductive behavior predicted by the model.

Finally, time-resolved measurements of electric polarization and current during cyclic voltage sweeps with varying amplitudes and periods will directly probe the predicted loop inversion phenomenon. These measurements will also facilitate the correlation of observed dynamics with the relaxation times associated with charge trapping and leakage processes.

The implementation of these experimental protocols will not only validate the theoretical model but also significantly enhance our understanding of complex dielectric systems and their potential applications in emerging devices.

\section{Conclusions}

In this work, we developed a comprehensive model to describe the dielectric response of non-polar, non-ferroelectric materials, focusing on the interplay between polarization transfer and charge leakage mechanisms. Our approach successfully reproduces key experimental phenomena, including the inversion of polarization loops and the emergence of ferroelectric-like hysteresis in leaky dielectrics. By differentiating true ferroelectric behavior from effects induced by non-equilibrium dipoles and energy-dependent transfer rates, the model provides a theoretical foundation for experimental observations and informs the accurate characterization of materials used in capacitors, transistors, and memory devices.

In capacitors, the model can guide the design of materials with reduced leakage and enhanced energy storage capabilities. For transistors, it can guide the development of high-performance FETs by optimizing gate dielectrics for improved control and reduced leakage. Furthermore, the model can contribute to the design of more reliable and efficient resistive RAMs and ferroelectric RAMs by providing insights into the factors influencing the stability and reliability of memristive and ferroelectric states.

Notably, the model highlights the tunability of dielectric responses through parameters such as voltage sweep rate and the dominance of polarization or leakage processes, offering insights into device optimization. Its ability to incorporate higher-order harmonics and multimode impedance analysis enhances its relevance for analyzing systems dominated by displacement currents.

Future extensions could explore material-specific adaptations, the effects of temperature on transport dynamics, and the role of external stimuli such as illumination and applied electro-magnetic fields. These advancements would further refine the model's applicability to diverse dielectric systems and its utility in guiding experimental protocols and the design of electronic devices.

\section{Appendix}
\begin{table}[h!]
\centering
\begin{tabular}{|c|c|c|c|c|}
\hline
\textbf{Figure} & $\tau^{(lk)}/\tau^{(p)}$ & $\lambda^{(lk)}/\lambda^{(p)}$ & $\eta$ & $\alpha$ \\ \hline
2(a) & -----/1.0 & ----- & $0.005$ & ----- \\ \hline
2(b) & -----/1.0 & ----- & $\sim$ 0.5 & 1.0 \\ \hline
3(a) & 1.0/0.1 & 10.0/100.0 & $0.01$ & 1.0 \\ \hline
3(b) & 1.0/0.1 & 10.0/100.0 & $0.01$ & 1.0 \\ \hline
4(a) & 1.0/0.1 & 10.0/100.0 & $0.01$ & 1.0 \\ \hline
4(e) & 1.0/0.1 & 10.0/50.0 & $0.01$ & 1.0 \\ \hline
5    & 1.0/0.1 & 1.0/2.0 & $0.01$ & 10.0 \\ \hline
6    & 1.0/0.1 & 1.0/2.0 & $0.01$ & 10.0 \\ \hline
7    & 1.0/0.1 & ----- & $0.01$ & 1.0 \\ \hline
\end{tabular}
\caption{Parameters used in each figure.}
\label{table}
\end{table}

\begin{table}[h!]
    \centering
    %\begin{tabular}{|c|c|}
    \begin{tabular}{|m{1.5cm}|m{6cm}|} 
    \hline
         $V$ & Applied voltage bias  \\\hline
         $\eta$ & Local voltage efficiency \\\hline
         $L$ & Length of the charge localization site \\\hline
         $P$ & Electric polarization \\\hline
         $\Omega = D\cdot A$ & System volume = system length $\cdot$ system area \\\hline
         $\partial q$ & Non-equilibrium charge fluctuation \\\hline
         $F^{(p)}$ & Non-equilibrium carrier flux contributing to space charge accumulation \\\hline
         $F^{(lk)}$ & Non-equilibrium carrier flux contributing to charge leakage\\\hline
         $\alpha$ & Symmetry control parameter \\\hline
         $T_{eff}$ & Effective temperature\\\hline
         $E^{(p)}$ & Effective barrier for carrier activation contributing to space charge accumulation\\\hline
         $E^{(lk)}$ & Effective barrier for carrier activation contributing to charge leakage\\\hline
         $\chi$ & Electric susceptibility \\\hline
         $E$ & Electric field \\\hline
         $\tau$ & Relaxation time \\\hline
         $j$ & Current density \\\hline
         $I$ & Current \\\hline
         $C$ & Geometric capacitance \\ \hline
         $G$ & Conductance \\\hline
         $B$ & Susceptance \\\hline
         $Z$ & Impedance \\\hline
         $Y$ & Admittance \\\hline
    \end{tabular}
    \caption{List of symbolsn}
    \label{tab:02}
\end{table}

\section*{Acknowledgments}
This study was financed in part by the Coordenação de Aperfeiçoamento de Pessoal de Nível Superior - Brazil (CAPES) and the Conselho Nacional de Desenvolvimento Científico e Tecnológico - Brazil (CNPq) Proj. 311536/2022-0. 
%\bibliographystyle{elsarticle-harv} 
%\bibliographystyle{elsarticle-num} % Estilo Elsevier Harvard
%\bibliography{ref2} % Nome do arquivo .bib (sem extensão)
%\bibliography{ref2.bib}
\bibliography{main.bbl} %arXiv

\end{document}